# X-RAY JET DYNAMICS IN A POLAR CORONAL HOLE REGION

BORIS FILIPPOV[1,3], LEON GOLUB[2] and SERGE KOUTCHMY[3]

[1]*Institute of Terrestrial Magnetism, Ionosphere and Radio Wave Propagation, Russian Academy of Sciences, Troitsk Moscow Region 142190 Russia*
(e-mail: bfilip@izmiran.troitsk.ru)
[2]*Harvard-Smithsonian Center for Astrophysics, 60 Garden Street MS58, Cambridge, MA 02138 United States*
(e-mail: golub@cfa.harvard.edu)
[3]*Institut d'Astrophysique de Paris, CNRS and Univ. P.& M. Curie, 98 Bis Boulevard Arago, F-75014 Paris, France*
(e-mail: koutchmy@iap.fr)

**Abstract.** New XRT observations of the north polar region taken from the X-ray Telescope (XRT) of the Hinode (Solar-B) spacecraft are used to analyze several time sequences showing small loop brightenings with a long ray above. We focus on the recorded transverse displacement of the jet and discuss scenarios to explain the main features of the events: the relationship with the expected surface magnetism, the rapid and sudden radial motion, and possibly the heating, based on the assumption that the jet occurs above a null point of the coronal magnetic field. We conclude that 3-D reconnection models are needed to explain the observational details of these events.

## 1. Introduction

A wide variety of jet-like structures are observed in the solar atmosphere. They can be formed both from relatively cool plasma as spicules, spikes, macro-spicules and surges (Beckers, 1972; Sterling, 2000; Yamauchi *et al.*, 2005; Rompolt and Svestka, 1996, Koutchmy and Stellmacher, 1976) as well as from hot plasma and seen as X-ray, white-light and EUV coronal jets (Brueckner, 1981; Shibata *et al.*, 1992; Shimojo *et al.*, 1996; Koutchmy *et al.*, 1997; Koutchmy *et al.*, 1998; Wang *et al.*, 1998; Chae *et al.*, 1999). The sizes of these structures, as measured by the lengths of the jets, varies from a few megameters (Mm) for spicules up to hundreds of Mm for large X-ray jets and to several solar radii for the fine linear rays seen in white light (Koutchmy and Nikoghossian, 2002). Of particular relevance to the numerous small events now being seen with XRT is the early work of Moore *et al.* (1977), relating EUV macrospicules to X-ray bright point flares (Golub *et al.*, 1974; 1977) and to small surge-like events seen in H$\alpha$.

There seems to be little doubt that all of these phenomena are collimated and guided by magnetic fields. However, although the mechanism of jet formation is not yet clear, many models try to ascribe to the magnetic field a more active role than merely guiding the mass flow. In extreme cases, the **J** × **B** force or electric fields have been suggested to accelerate the plasma flows (see Shibata and Uchida, 1986; Lorrain and Koutchmy, 1996; Henoux and Somov, 1997). Schluter (1957) proposed a 'melon-seed' mechanism for diamagnetic plasma to be forced out of the diverging magnetic field. Platov, Somov, and Syrovatskii (1973) calculated the raking-up of plasma due to the growth of the local magnetic field. For a surge formation in ~ 1 min, the magnetic moment of a small magnetic concentration (local dipole) should grow by a factor of ~25, which presently seems too fast for observed changes in photospheric fields. Filippov (1993) examined surge formation by the compression of the lower part of a vertical magnetic tube owing to the growth of the local magnetic field. A system of hydrodynamic equations for an ideal gas in a flux tube was solved numerically. An increase in the field strength at the photospheric end of the vertical flux tube with a time scale of less than 300s is necessary for the efficient formation of a jet having the characteristics of a typical surge.

Much attention has been paid to spicule models in which field line reconnection takes place, starting from Pikel'ner (1969), Uchida (1969), Kosovichev and Popov (1978), Blake and Sturrock (1985) and going to the more recent numerical simulations by Shibata *et al.* (1994), Karpen, Antiochos, and DeVore (1995), Yokoyama and Shibata (1996), Karpen *et al.* (1998). In this paper, we analyze new Hinode X-ray observations taken over the northern polar coronal hole region. Numerous events characterized by loop brightenings with overlying jet-like phenomena were observed (Cirtain *et al*., 2007; Culhane *et al*., 2007; Savcheva *et al*., 2007; Shimojo *et al*., 2007). They are quite similar to larger-scale and less often observed jets in X-ray and EUV with longer lifetime and occurring inside active regions as originally reported by Shibata *et al*. (1992; 1994) based on Yohkoh SXR images. Jets were episodically seen in EUV over polar regions with the EIT (SoHO) 195Å channel when using fast partial frame images but no detailed analysis were apparently published. With the XRT these events are found to be more frequent, and greater structural detail is seen thanks to the improved temporal and spatial resolution of this instrument. This improved observational capability permits a more rigorous confrontation with the theoretical models. The purpose of this paper is to examine the dynamics of such events in a 3-D configuration. We propose that the simple 2-D reconnection picture does not apply and that at least the acceleration phase of these events can proceed without magnetic reconnection energy release as it is usually understood from 2-D models.

## 2. Observational data

We selected for this analysis two periods of observation of the North polar region with the X-ray telescope (XRT) on board the Hinode (formerly Solar-B) Observatory. The first period begins from 00:47 UT till 07:44 UT on 23 November 2006, and the second one, from 16:09 until 22:00 UT on 9 January 2007. The XRT (Golub *et al*., 2007) provides about 1-arcsecond resolution images of the solar corona with temperatures from ~1 MK to ~20 MK (6--300Å). The cadence of images on both occasions was 30 seconds.

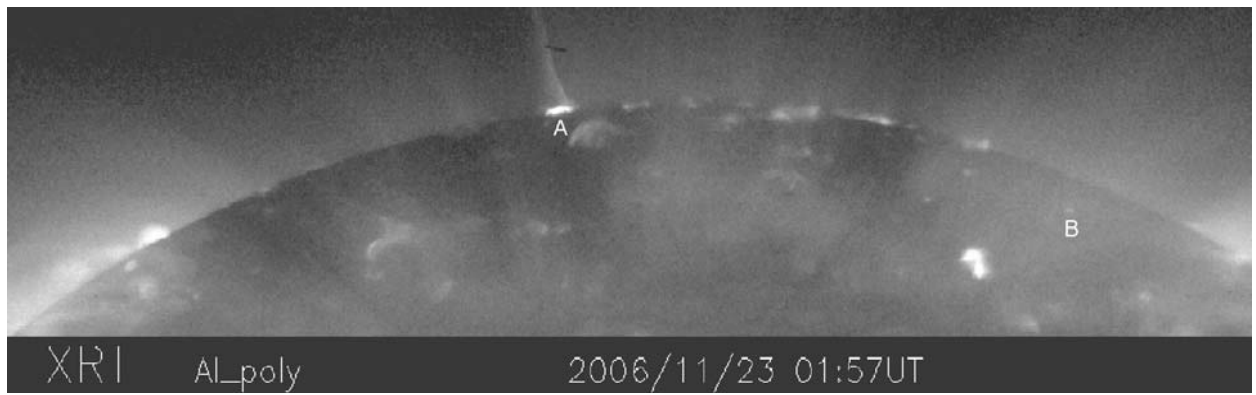

*Figure 1*. X-ray image of the solar northern polar corona, obtained with Hinode XRT on 23 November 2006.

Figure 1 shows an example of an X-ray coronal image. Since solar cycle 23 at the end of 2006 was not far from the minimum of activity, the northern polar region was covered by a polar coronal hole, which appears as a dark area in the X-ray images. However a number of small bright formations which have been called X-ray bright points in the past (Golub *et al.,* 1974) are visible within it. The smallest brightenings are not fully resolved and they show up as bright specks, while larger structures can be recognized as loops or loop systems. From time to time certain formations become brighter or suddenly appear in the dark background. The process lasts from several minutes to half an hour and may reappear after some time, almost at the same location. As a rule, a small jet-like structure appears above the brightening. We will refer to this type of event as a jetlet rather than extend the name "jet" as others have done. During 7 hours of observations on 23 November 2006, at least 24 different jetlet events were counted within the field of view shown in Figure 1.

There would seem to be little doubt that jetlets are elongated along the direction of the large-scale surrounding magnetic field. Their axes are nearly straight when they originate from sites in the middle of a coronal hole (CH) near the pole (Figure 1) where magnetic field lines should be straight. They are curved (Figure 3) when they originate from sites near the boundary of the coronal hole where field lines are curved due to an apparent super-radial divergence.

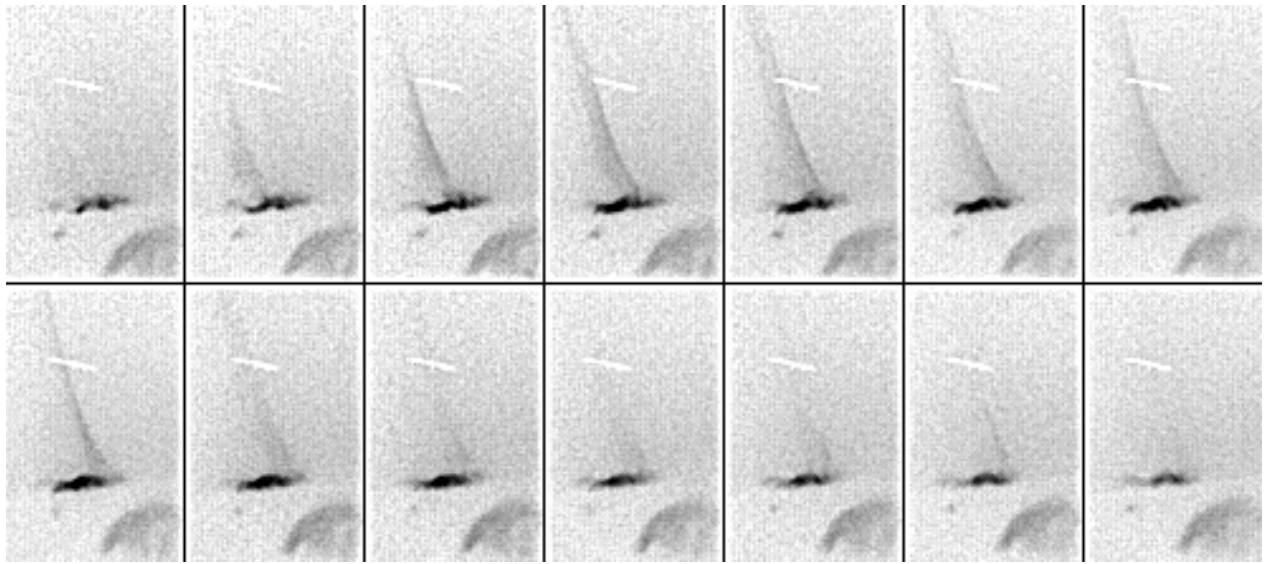

(a)

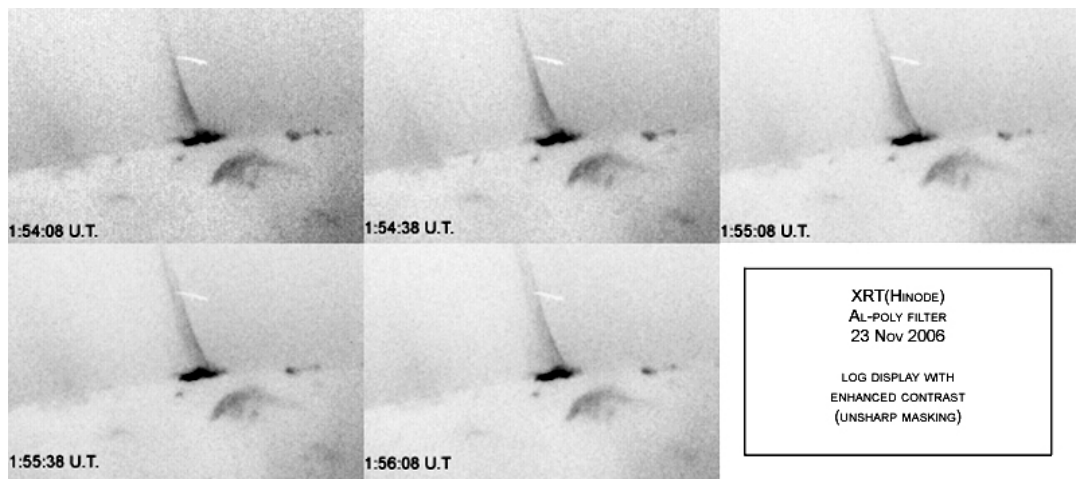

(b)

*Figure 2.* Negative X-ray images showing the jetlet evolution between 01:53 UT and 02:06 UT on 23 November 2006 (a). This jetlet is visible in Figure 1, labeled A. Time intervals between individual frames are 1 min. The white segment at the top of each image is produced by a speck of dust in front of the CCD detector. (b) Detailed view with 30 sec cadence of the apparent transverse motion in the jet between 01:54:08 UT and 01:56:08 UT. The high time cadence shows that there are two successive jets formed alongside each other, leading to an appearance of transverse motion in a single jet.

The region designated as A in Figure 1 was notably active. Five jetlets were observed within a 7-hour period of observation on 23 November 2006. The development of the second of these jetlets is shown in Figure 4. Jetlets were then observed within the intervals 04:29 UT – 04:40 UT, 04:47 UT – 05:04 UT, and 06:50 UT – 07:07 UT. The region was obviously located partly beyond the limb, however very close to the visible hemisphere, so that only the upper parts of the bright loops lying at the base of the jetlets was visible. The region is also inside the N-pole CH.

With time the region disappeared completely. It is not unlikely that jetlets within a polar coronal hole coincide with macrospicules.

A jetlet appears first as a rather faint structure, with the brightness increasing. It is difficult to measure the speed of the upward motion: due to the high temperature the scale height is a significant fraction of the length of a jetlet and one can barely distinguish the leading edge of the jetlet. A faintly visible upper boundary of a jetlet is seen in the first four frames in Figure 5 and the average velocity derived from these images is about 400-450 km s$^{-1}$. A statistical analysis performed by Savcheva *et al.* (2007) using 104 of these events selected inside the S-pole CH shows that the most probable speed of the projected longitudinal motion is 160 km/s with extreme values reaching 500 km s$^{-1}$. The same work provides speeds of the projected *transverse* motions from 0 to 40 km s$^{-1}$.

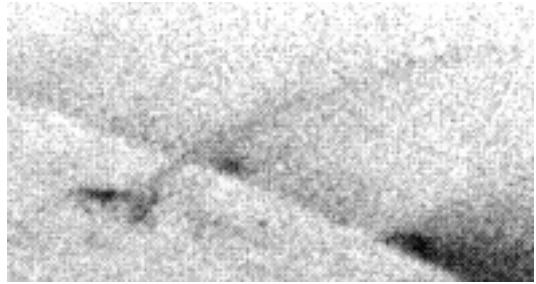

*Figure 3*. Negative X-ray image of a jetlet at 04:22 UT on 23 November 2006. The location of the source region of the jetlet is labeled B in Figure 1.

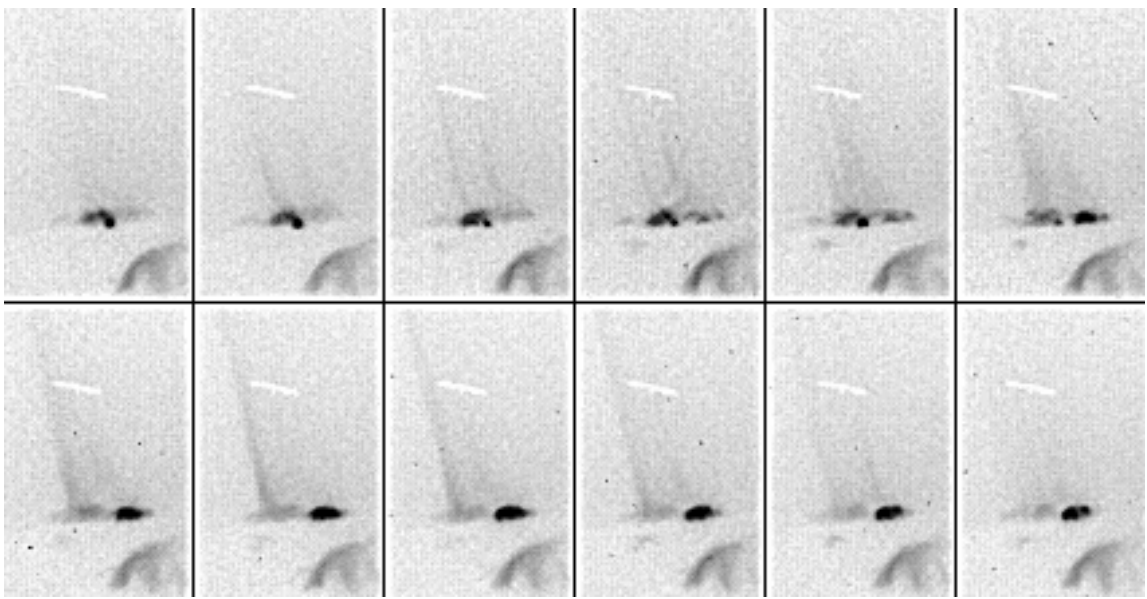

*Figure 4*. Negative X-ray images showing the second jetlet in the region A between 03:14 UT and 03:25 UT on 23 November 2006. Time intervals between these frames are 1 min, although images were acquired every 30 sec. Note that the growth of the jetlet at left continues after the footpoint brightening.

What is curious is that many jetlets show a rather fast lateral or transverse motion. It is clearly noticeable in Figure 2 and Figure 4. (A speck of dust on the CCD detector can serve as a fiducial mark, which helps to precisely measure the lateral displacements.) The speed of horizontal motion is about 25 km s$^{-1}$ and it looks as if the entire structure moves transversely. However, it is difficult to decide whether we see a definite real horizontal motion of the structure or whether we see an apparent change due to the displacement of the source region of the flow; the latter possibility does sometimes seem to occur (*viz.* Figure 2a). The length of the jetlet is about 60

Mm. With a vertical velocity estimated to be ~400 km s$^{-1}$, the plasma can travel nearly half of the jetlet's length in one minute. The transverse displacement occurs over 5-10 minutes, so that it is not physically the same plasma that is moving transversely.

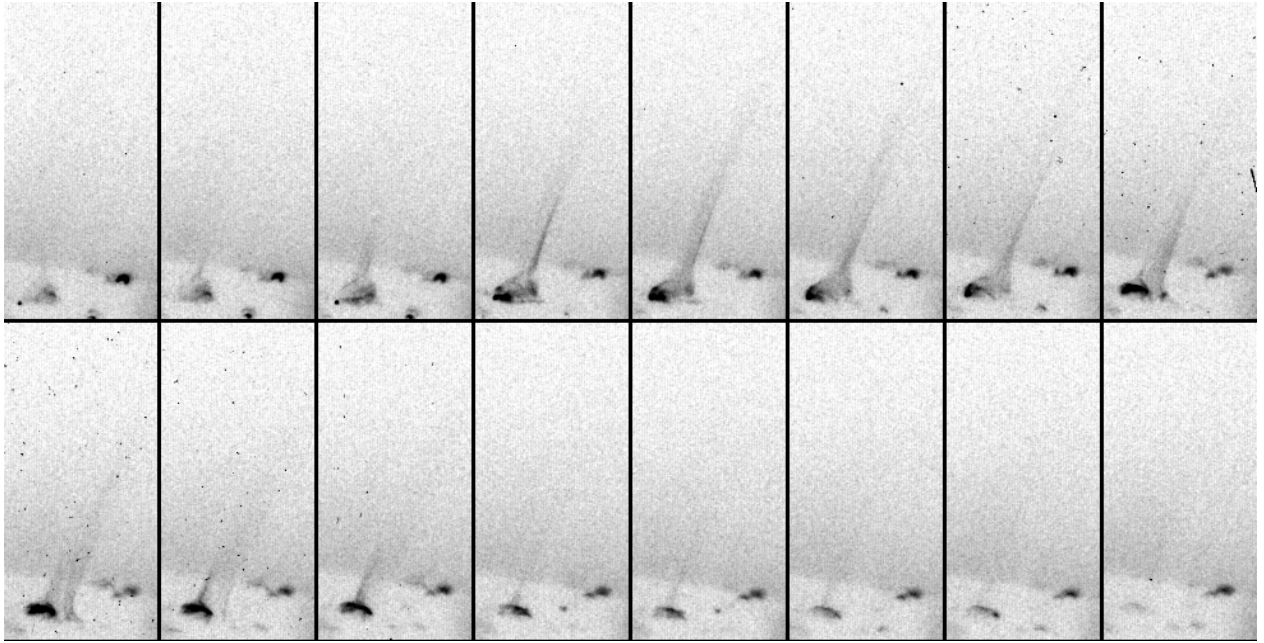

*Figure 5.* Negative X-ray images showing the jetlet evolution on 09 January 2007. Time intervals between frames are 1 min.

### 3. Jetlets and the structure of the magnetic field

As mentioned above, there is no doubt that the shape of jetlets is related to the magnetic field. From their shape, we can infer not only the general direction of the symmetry axis of the structure but also its exterior appearance. The jetlets resemble miniature helmet-streamers: they are relatively wide at the base and then become increasingly narrow with concave boundaries. The jetlet's shape resembles the geometry of field lines in the vicinity of a null point (or a neutral point, or an X-type singular point). This is not an accidental analogy because at the base of practically every jetlet we see a loop or a system of loops whose superposition with the nearly vertical and homogeneous magnetic field of a coronal hole requires the existence of null points.

Unfortunately, magnetic field measurements in polar regions cannot provide enough sensitivity and precision for a meaningful comparison with the observed topology, and one cannot find the counterparts to these small loops in magnetic field maps. Nevertheless, there is no doubt that all of them are related to either small patches of parasitic polarity or small bipolar regions.

These two cases can be schematically represented by a vertical or horizontal bipole inside a large unipolar region (Shibata *et al.*, 1994; Shibata *et al.*, 1996; Canfield *et al.*, 1996; Filippov, 1999). Figure 6 shows a 2-D cut through the field lines of the potential (current-free) field for the two possibilities. The potential field seems reasonable for the low corona where magnetic pressure is much greater than gas pressure due to the very low densities inside the CH. For a 2D representation, these schemes correctly show the topology of field lines in the plane passing through the null point and the dipole. For the vertical dipole, the 3D pattern can be visualized by making an axial rotation of Figure 6b around the central vertical axis. For the horizontal dipole, the situation is more complex and visualization is difficult. Field lines are flat only in a single plane of symmetry. Any other cross-section of the magnetic field containing the null point would have magnetic field components perpendicular to the cross-sectional plane.

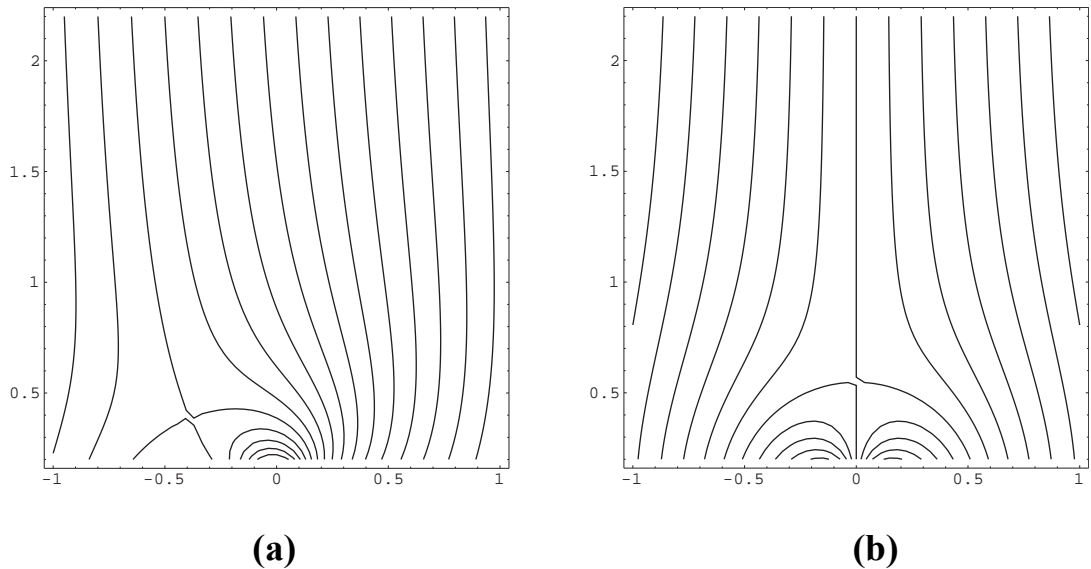

**(a)**          **(b)**

*Figure 6.* Model of magnetic field emergence into a unipolar background: parasitic unipolar field (a) or bipolar field (b). Magnetic field lines of a 2D dipole are superposed on the surrounding vertical homogeneous field. The dipole axis can be horizontal (left) or vertical (right).

    The magnetic field configuration shown in Figure 6b was clearly observed in the emission line corona at larger scales using SOHO/EIT coronal images (Filippov, 1999). Below the central part of the structure, there was a patch of parasitic polarity within a large-scale unipolar area. Shibata *et al.* (1994) described a very strong X-ray jet formation in such a configuration using the Yohkoh observations, calling such a structure an anemone active region. A quite similar dome-like configuration in 2.5-D was suggested by Koutchmy and Molodensky (1993) to explain the observation of white light active region coronal rays described by Koutchmy (1969). The configuration with a horizontal dipole (Figure 6a) was found in TRACE images taken on the disk on 3 October 2001 (Filippov, Koutchmy, and Vilinga, 2007). The growth of the dipole was well documented using the SOHO/MDI magnetograms, and field aligned motion above the null point in the corona was attributed to the process of emergence of the magnetic dipole.

    While the shape of the jetlets seems to be controlled by a magnetic field similar to the ones shown in Figure 6, the origin of the driving force of the flow remains unclear. As bright loops are observed at the feet of every jetlet, the pressure gradient caused by the high temperature is a plausible force that causes plasma to move. But what is the heating mechanism? The presence of the neutral point naturally suggests that field line reconnection and magnetic field annihilation could occur in it. Shibata *et al.* (1994) came to the conclusion that a magnetic reconnection mechanism is responsible for the large X-ray jet production.

    It is important to bear in mind that the 2-D models of reconnection are not applicable when a 3-D geometry is obviously needed for the jetlets. There is no place where a classical current sheet could develop, since there are no separators in this model. (Recall that a separator is a field line that connects two null points and there is only one null point in the configuration we are considering.) For the configuration of Figure 6b, no reconnection takes place during the growth or the fading of the dipole if the axial symmetry is conserved. No field line crosses the null point in this case. During the growth of the horizontal dipole, the nearly vertical field lines of the unipolar region rearrange, moving from the left side of the dipole to the right side. Some field lines pass through the reconnection in the X-point while most of them are able simply to flow

around the null point (fan reconnection), which is impossible to represent in a 2-D geometry (Priest and Titov, 1996). In addition, observations do not reveal any local emission near the place of the null point but bright loops appear below the separatrix surface. This fact was also noticed by Shibata *et al.* (1994); they suggested that the released energy near the X-point was very quickly transferred into the reconnected lower loop and the reconnected open field. These two difficulties – the lack of observed heating at the supposed reconnection site and the ability of field lines to evade reconnection in a 3-D geometry – lead us to search for an alternative scenario to explain the observed jetlet events.

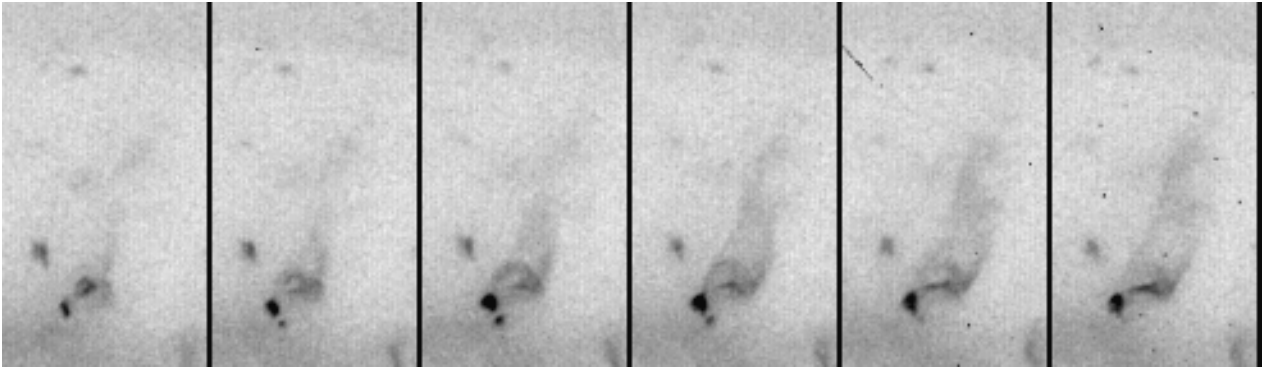

*Figure 7.* Negative X-ray images showing an expanding loop between 20:11 UT and 20:16 UT on 09 January 2007. Time intervals between frames are 1 min.

## 4. Jetlets not initiated by reconnection energy release

One possible scenario is that the emerging loop, with its enhanced gas pressure grows into the pre-existing field. As the emerging loop approaches the null point, the local field strength decreases and the gas pressure exceeds the magnetic pressure, so that the plasma is able to expand. The easiest way for expansion to proceed is through the null point, where the magnetic field is weak: then a strong flow will rapidly appear above the closed loop system. The strong vertical magnetic field above the null point collimates the plasma flow as a nozzle. If the gas pressure is high enough and given the extra degrees of freedom in the 3-D field topology compared to the 2-D case, the expanding loop can force the field-lines apart, as is possibly seen in Figure 7.

A version of this scenario is illustrated in Figure 8, which shows the ejection of an individual twisted flux tube (small flux rope) within the dome-like magnetic configuration. If new emerging magnetic flux is twisted enough and in addition non-homogeneously, a portion of the tube with the largest twist will lose equilibrium and erupt, as do so large-scale flux ropes during filament eruptions and CME onsets. We can consider that a separate twisted flux tube emerges from below the photosphere and grows into the nearly potential magnetic field configuration with a null point. Depending on its free magnetic energy, i.e. the strength of the electric current in the tube, the flux tube will evolve quasi-statically or dynamically.

If the free (non-potential) energy density is small compared to the potential magnetic energy density, the twisted flux tube will evolve quasi-statically. During its ascent within the dome of closed field lines its shape will be approximately that of the shape of the surrounding potential field lines, and in the same way as the other tubes it can pass through the reconnection point. However, the twisted flux tube can contain hot plasma heated during the emergence process, as is often observed with emerging flux. Hot plasma confined in the magnetic trap of the closed flux tube can now spread along the open field lines after reconnection and form a jet along the magnetic field (Figure 8d).

The eruption of several individual rising tubes – within this twisted flux rope can create the pattern of subsequent closely-spaced neighboring imitating the lateral motion of a single jetlet.

Note that plasma was heated within the emerged twisted flux before the reconnection and the presence of the null point only allowed it to spread along the open field near the null point. The role of the twist here is as the source of free energy for plasma heating and then for the rise of the individual tube within the dome-like magnetic configuration.

If the stored free energy (electric current) is large, the process can proceed violently after the sudden loss of equilibrium as happens in models of large-scale flux ropes (Forbes and Priest, 1995; Lin *et al.*, 1998). The expansion and ascent of the twisted loop will not be influenced by weak field in the vicinity of the null point, but later and at larger heights stronger magnetic field is able to guide plasma motion (Figure 8c). In both cases the external field, having the shape of an inverted funnel, will collimate and guide the upward flow.

The role of the magnetic field can be more active than just to guide the mass flow, if the field changes rapidly enough. As seen in Figure 6, the field lines are denser on one side of the horizontal dipole and at both sides of the vertical dipole. The field line density increases with the increase of the dipole momentum. This means that the cross-section of an individual flux tube decreases at this place, which leads to an increase of the gas pressure. So a pressure growth at the base of some flux tubes would initiate field aligned motion of the coronal plasma.

The frozen-in plasma motion in the vicinity of a null point also creates some pressure inhomogeneities that are able to cause a field aligned motion. In a first approximation the trajectories of a frozen-in plasma flow are orthogonal to field lines in each point. The hyperbolic shape of the field lines leads to a plasma compression within two diametrically opposed quadrants, where the flows converge, and a rarefaction within the two other quadrants, where the flows diverge (Somov and Syrovatskii, 1971; Filippov, 1997; Filippov *et al.*, 2007). The action of this geometrical factor is enhanced by the effect of acceleration of the plasma approaching the null point as $v \sim E/B$ and $B$ is decreasing (div $\mathbf{v} > 0$). The plasma that outflows from the null point decelerates (div $\mathbf{v} < 0$). So, by the equation of continuity

$$\frac{\partial \rho}{\partial t} = -\mathrm{div}\rho\mathbf{v} ,$$

the regions of enhanced density (and enhanced pressure) should appear right-and-above the center of the saddle structure in Figure 6a and left-and-down of it during the growth of the dipole.

The remaining question is: does the plasma flow along the visible structure for the entire duration of the visible event or does it rise into the vertical field and then stay in a quasi-static state as is thought to be the case in coronal loops? In the former case, the lateral motion suggests the displacement of the flow source region. In the latter case, the lateral motion means that there are changes in the magnetic field that allow the displacement of field lines up to a rather large height. This is a major issue that could be resolved if a time sequence of resolved magnetograms were taken above the polar regions. Numerical 3-D simulations of the behavior of the coronal plasma around the null point are another way to proceed to resolve this issue.

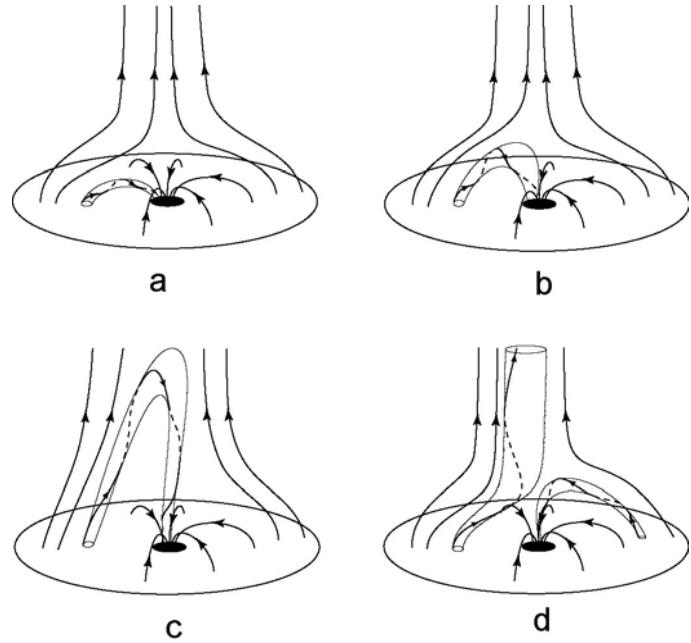

*Figure 8.* Scheme of possible scenario of ejection of a small twisted flux tube within dome-like magnetic configuration. (a) and (b) show the growth of an emerging bipole into the assumed unipolar field; (c) and (d) show two alternate possibilities. In (c) the loop moves upward through the null point region, while in (d) it reconnects with the ambient field and one leg opens out.

## 5. Conclusions

Polar coronal holes show a significant activity manifested in small loop brightenings and jetlet formation. Geometrical shape of the jetlets and their position indicate that they appear near the singular points of the magnetic field, namely, null points or X-points. These nulls arise due to the interaction between new emerging small dipoles and large-scale magnetic field of the coronal hole. Observations show that a jetlet event starts from the brightening and expansion of a small loop within a dome-like magnetic configuration. We discuss several scenarios of a jetlet formation alternative to the well-known model of Shibata *et al.* (1992; 1994) call a reconnection model within an "anemone-type" configuration. The latter is indeed based on 2-D numerical simulations which may not adequately describe the real 3-D pattern of the event. At least no manifestations of energy release are observed in the immediate vicinity of the null point, while brightening starts at low altitude and then spreads upwards, ending in jetlet formation.

We propose a qualitative model of jetlet formation within the "dome-like" magnetic configuration in which the energy source of the process is not energy release in a reconnection site but free magnetic energy of a small twisted flux tube. Plasma is heated in this tube and confined in a magnetic trap. Coming to the reconnection site near a null point, where the magnetic field is weak, the trap is emptied and hot plasma is able to rapidly move upward along open field lines. Gas pressure gradient works like a piston that pushes plasma through a nozzle. If the electric current within the twisted flux tube is strong enough, the flux tube erupts more violently as a small CME.

The model is conceptual and needs to be confirmed by numerical calculations. We realize that this task is not easy because of the 3-D configuration and inclusion of a null point. The mechanism of 3-D reconnection is unclear and only recently theoreticians have begun to study it. A full model should include description of the 3D flux rope evolution and peculiarities of the 3-D reconnection.

The jetlets reveal some real observable vertical mass motion from the CH region. It is difficult to measure the radial velocity of plasma within the jetlets but, given the large number of such events now being seen by the XRT, it seems large enough to significantly contribute as a large momentum input to the solar wind. The modeling needed to quantify this contribution to the fast solar wind is currently being undertaken as part of the continued studies of these dynamic coronal features.

## Acknowledgements

This work was supported in part by the Russian Foundation for Basic Research (grant 06-02-16424) and in part by the NATO CRG 940291. Hinode is a Japanese mission developed and launched by ISAS/JAXA, with NAOJ as domestic partner and NASA and STFC (UK) as international partners. It is operated by these agencies in co-operation with ESA and NSC (Norway). US members of the XRT team are supported by NASA contract NNM07AA02C to SAO. The work is partly completed thanks to exchange visits to the Paris Institut d'Astrophysique.